# Statistical Information of the Increased Demand for Watch the VOD with the Increased Sophistication in the Mobile Devices, Communications and Internet Penetration in Asia

Saleh Ali Alomari[1] and Putra Sumari[2]

[1,2]School of Computer Science, Universiti Sains Malaysia
11800 Pulau Penang, Malaysia
salehalomari2005@yahoo.com and putras@cs.usm.com

## ABSTRACT

*As the rapid progress of the media streaming applications such as video streaming can be classified into two types of streaming, Live video streaming, Video on Demand (VoD). Live video streaming is a service which allows the clients to watch many TV channels over the internet and the clients able to use one operation to perform is to switch the channels. Video on Demand (VoD) is one of the most important applications for the internet of the future and has become an interactive multimedia service which allows the users to start watching the video of their choice at anytime and anywhere, especially after the rapid deployment of the wireless networks and mobile devices. In this paper provide statistical information about the Internet, communications and mobile devices etc. This has led to an increased demand for the development, communication and computational powers of many of the mobile wireless subscribers/mobile devices such as laptops, PDAs, smart phones and notebook. These techniques are utilized to obtain a video on demand service with higher resolution and quality. Another objective in this paper is to see Malaysia ranked as a fully developed country by the year 2020.*

## 1. INTRODUCTION

As a new eradicating application in the recent few years in video content generation and distribution the Internet has observed a great expansion of networked video sharing. This is indicated from a large number of sites that make it easy for users to upload and share videos. Certain sites such as YouTube [1] which is the most successful and the largest video sharing Web site on the Internet with over 100 million video being watched every day and 65,000 video being uploaded per day [2], furthermore it is the most trafficed sites on the web.

According to a report from YouTube in June 2010, more than 177 million Internet users watched video contents in June. In the Google Sites, which are driven primarily from video viewing at YouTube and become the top video content property almost 144.5 million unique users, as well as Yahoo which has almost 44.9 million users and Vevo which has almost 43.7 million users. The highest number of overall viewing sessions with 1.8 billion and average time spent per viewer at 261 minutes, or 4.3 hours was observed with Google Site. Furthermore, with an average of 135 minutes (or 2.2 hours) per viewer was Hulu [3]. Additionally, social networking websites such as

                                                                    

MySpace [4] have brought many of the features which help to make MySpace playing an important role in our life and become a part of daily life. A new blog subscription service has been introduced where one can join a favourite blog and enjoy all its content without ever having to leave MySpace. Furthermore, FaceBook [5] being the most popular website which is used by millions of people every day, and is accessed by more than 150 million active users currently through mobile devices to upload unlimited number of photos, share links and videos. Further details about facebook [5] are shown in Table 1.

Table 1. Statistics of the Facebook [5]

| | |
|---|---|
| **People on Facebook** | • More than 500 million active users<br>• 50% of our active users log on to Facebook in any given day<br>• Average user has 130 friends<br>• People spend over 700 billion minutes per month on Facebook |
| **Activity on Facebook** | • There are over 900 million objects that people interact with (pages, groups, events and community pages)<br>• Average user is connected to 80 community pages, groups and events<br>• Average user creates 90 pieces of content each month<br>• More than 30 billion pieces of content (web links, news stories, blog posts, notes, photo albums, etc.) shared each month. |
| **Global Reach** | • More than 70 translations available on the site<br>• About 70% of Facebook users are outside the United States<br>• Over 300,000 users helped translate the site through the translations application |
| **Platform** | • More than one million developers and entrepreneurs from more than 180 countries<br>• Every month, more than 70% of Facebook users engage with Platform applications<br>• More than 550,000 active applications currently on Facebook Platform<br>• More than one million websites have integrated with Facebook Platform<br>• More than 150 million people engage with Facebook on external websites every month<br>• Two-thirds of comScore's U.S. Top 100 websites and half of comScore's Global Top 100 websites have integrated with Facebook. [19] |
| **Mobile** | • There are more than 150 million active users currently accessing Facebook through their mobile devices.<br>• People that use Facebook on their mobile devices are twice more active on Facebook than non-mobile users.<br>• There are more than 200 mobile operators in 60 countries working to deploy and promote Facebook mobile products |

The wide spread of wireline broadband Internet access, cable and DSL, with which it becomes much faster to download/upload high resolution videos are some of the factors that lead to technological advances. In addition, the portable devices with cameras being ubiquitous, such as smart phones and PDAs, facilitate regular users to conveniently create videos at a reasonable quality instantaneously and videos can then be transmitted to each other through their mobile devices, services as such lead to increase the demand of users and development of latest mobile appliances with smooth features to watch videos in high quality. Most importantly the mobile devices allow surfing on the internet and specially users' mobile devices remain connected 24 hour to the social websites such as Facebook and MySpace for feed back messages and user can reply via internet service on mobile. According to the latest research from Analysis, revenue from mobile media and entertainment (MME) [7] [18] services in the US will be doubled during the period of next five years. Revenue of mobile media and entertainment services US generated US$3.1 billion in 2007, and analysis from the research forecasts that revenue growth rate can increase to $6.6 billion in 2012 [9], at a compound annual growth rate of 16.3%. After 2010 growth rate will not increase unless the technical and market environment for MME services improve, forecasts according to the latest analysis report on mobile media and entertainment MME services in the US during the period of 2007-2012. Revenue of non-voice service will account for 12.3%, forecasts according to the research analysis on MME services in the US by the year of 2012. Highest growth rate will be experience by the Mobile TV and VoD services of any MME service during the next five years [23][24]. Broadcast, multicast and unicast TV and video services combined revenue will account for 36% of MME by 2012.

In contrast, revenue from personalization services will decline from 47% of total MME revenue in 2007 to 17% in 2012 [6]. A recent study by Ground Truth, a mobile measurement firm, revealed that approximately 60% of the time spent on the mobile Internet is spent on social networking sites and apps. The second most popular category is portals on which Users spent only about 14% of the time on mobile Internet. The disparity of time spent between both category social networking and portals, account for 59.83 % and 13.65 % percent of time spent respectively, it is a vivid illustration of the impact that social networking has on Mobile Internet traffic in a week is shown in the fgure 1[8].

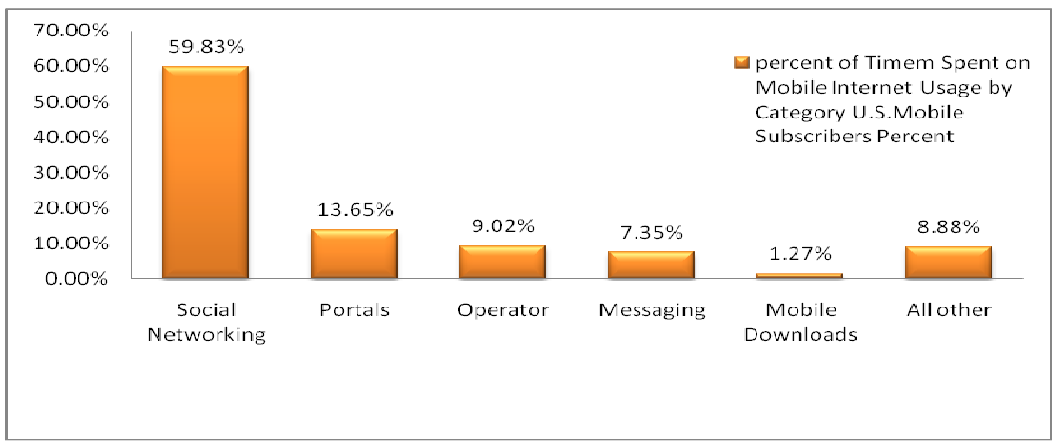

*Source: Ground Truth, Inc. Census of mobile subscribers for the week ending April 4, 2010 (n=3.05 Million U.S. Mobile subscribers)*

Figure1. Shows the mobile Internet traffic in a week at end of 2010

As shown in the Figure 1 the statistical data was calculated over two weeks period in the mid od 2010, and its shows that more than 3.05 million mobile phone users in the America. As well as the study revealed to the Social Networking Platforms (SNP) designed for the mobile devices, such as AirG, MocoSpace, Mbuzzy, MobaMingle, MobaMingle and etc, all of this SNP have a higher engagement levels compared to the platforms geared primarily for Personal Computer (PC) access such as ( Facebook and MySpace). As shown in the Table 2, we classified the mobile SNP Usage USA Mobile Subscribers week in the mid of 2010, the mobile users of MocoSpace and AirG spent almost three times as many minutes on those platforms than users in the Facebook, as well as more than double the amount of time mobile users interacted with MySpace on their mobile devices and the 2ed is the Facebook.

Table 2: Mobile Social network Usage U.S. Mobile Subscribers week ending April 4, 2010

|  | Sessions per Subscriber | Pages Per Subscriber | Page Per Session | Time Per Subscriber |
| --- | --- | --- | --- | --- |
| Average | 68.1 | 310 | 4.56 | 0:52:12 |
| Myspace | 57.6 | 246 | 4.28 | 0:40:41 |
| Facebook | 56.9 | 205 | 3.61 | 0:30:54 |
| Mocospace | 63.9 | 476 | 7.45 | 1:31:02 |
| FunForMobile | 17.4 | 101 | 5.83 | 0:19:50 |
| AirG | 5.8.8 | 520 | 8.84 | 1:13:03 |
| Facebook Photos | 18.9 | 59.7 | 3.15 | 0:10:10 |
| MobaMingle | 13.5 | 145 | 10.8 | 0:23:55 |
| Mbuzzy | 64.3 | 359 | 5.58 | 1:09:41 |
| Mocospace Photos | 15.7 | 57.2 | 3.63 | 0:12:22 |
| MobaMingle | 42 | 278 | 6.62 | 0:47:06 |

*Source: Ground Truth, Inc. Census of mobile subscribers for the week ending April 4, 2010 (n=3.05 Million U.S. Mobile subscribers)*

The latter suggests that there are opportunities for much greater engagement on both Facebook and MySpace via mobile. AirG and MocoSpace are chat-based platforms. Facebook and MySpace might want to look into including greater chat functionality with their apps, given the amount of time users spend engaging with those services each week.  By distributing high-speed Internet access from cable, Digital Subscriber Line (DSL), and other fixed broadband connections but within wireless hotspots, WiFi has dramatically increased productivity and convenience.

Currently, the WiFi delivers high speed Wireless Local Area Network (WLAN) connectivity to millions of clients over the world such as in their homes, offices, hotels, cafes, restaurants, airports and etc. In the worldwide, almost more than 223 million homes have WiFi connections, and there more than 127 million WiFi hotspots connections [10] [11] [14]. The integration of the WiFi connections into notebooks has accelerated the adoption of WiFi connections to the point where it is nearly a default feature in notebooks and at any mobile devices such as (PDAs, Laptops, mobile phone and etc.). In this care, more than 97% of laptops ship with WiFi integrated [2] and that lead to increase the number of handhelds and Consumer Electronics (CE) devices are adding WiFi capabilities. In the WiMAX takes wireless Internet access to the next generation, and with the time could reach the similar attach rates to devices as WiFi connections. WiMAX can provide Internet access miles from the nearest WiFi hotspot and blanket large areas. In the Wide Area Networks (WANs) can be they metropolitan, suburban, or even can be a rural with multi megabit per second mobile broadband Internet access [3]. In other hand the wide area Internet connectivity offered at the 2.5 Generation and 3ed Generation (3G) cellular data services has been mobile devices, and these services do not provide the broadband speeds to which users have become accustomed and that WiMAX can deliver. Since few years ago the WiMAX has established its relevance as an alternative to wired such as Digital Subscriber Line (DSL) and the normal cable providing a competitive broadband service offering that can be faster and cost effectively deployed. In this time the mobile WiMAX is defined in the Institute of Electrical and Electronic Engineers (IEEE 802.16e) standard was developed in 2005 and adds broadband connectivity on the move. And mobile WiMAX based on the scalable Orthogonal Frequency Division Multiple Access (OFDMA) technology [7] and this technology is capable of simultaneously supporting fixed, mobile usage models and portable. Furthermore, with scalable OFDMA, the operators no need a long time to choose between fixed or mobile services. In the both techniques ( WiMAX and WiFi) are ideal partners for service providers to deliver convenient, affordable mobile broadband Internet services in more places. Also both are based on IEEE wireless standards built from the ground up for Internet Protocol (IP) based applications and services. By combining between the WiMAX and WiFi access together, the service providers can deliver high speed Internet connectivity that subscriber's desire in more places. Finally, the WiMAX and WiFi technology synergies enable seamless integration into many of the mobile devices such as PDAS, laptops, mobile phone, CE devices, and its become exist at any of the new category of devices called "mobile Internet devices. In the Table 3 shown more details about the both techniques Wifi (IEEE 802.11 a/g/n) and WiMAX (IEEE 802.16e) [12] [15].

Table 3. Comparison between Wifi and WiMAX

| WiFi | WiMax |
|---|---|
| IEEE 802.11 a/g/n | IEEE 802.16e |
| Deployed in local coverage areas, such as public hotspots, homes, and businesses. | It's deployed in wide coverage areas, including metropolitan areas for mobile broadband wireless. Furthermore, the rural or remote areas for last mile connectivity and portable service. |
| Products certified by the WiFi Alliance | Products certified by the WiMAX Forum |
| Embedded in 97% of laptops and many handheld and CE devices | Customer Premise Equipment (CPE) and PC cards available today; embedded in laptops and handheld devices starting in 2008. |
| Provides fixed and portable solutions | Provides fixed and portable solutions |
| Operates in license-exempt spectrum. Current solutions use the 2.4 and 5 GHz bands | Operates in licensed spectrum and the Current solutions use the 2.3 GHz, 2.5 GHz, and 3.5 GHz. |
| It's cover a short range within 100 meters for each single access point. | The coverage area of the metropolitan up to several kilometres for a single BS. (fixed and lower density deployments) |
| At any devices connect through a wiFi Access Point (AP) to the operator's IP network and to the Internet connection. | At any devices connect through a Base Station (BS) to the operator's IP network and to the Internet connection. |

| It's evolution to mesh networks in metropolitan areas and its can cover a wide area. | It's evolution to multihop relay to improve range, data rates and increase the services area. |
|---|---|
| AP that include WiFi for access and WiMAX for network connectivity. | In the leverages digital advances so that the entire (BS) can be mounted on tower tops. |
| Channel access .11./11b: DSSS, .11a/g OFDM | LOS: SC, NLOS: SCa, OFDM, OFDMA .16e:OFDMA |
| Duplexing Done by MAC | TDD or FDD with support of MAC |
| The channel bandwidth is 20 or 22 MHz | LOS: 25 or 28 MHz, NLOS: 1.25-28 MHz |
| Max. velocity of MT is 30 km/hr | .16e: 70 km/hr |

## 1.2 BROADBAND CAMPUS COVERAGE

Many enterprise, government, and educational organizations have deployed WiFi in buildings for their work force and students. In the new technology of WiMAX which provide a services to offer broadband connectivity outdoor environment to provide blanket coverage of an entire campus such as (outside the buildings). The integration of WiMAX and WiFi onto a common device platform enables users to connect to either in-building WiFi or campus-wide WiMAX networks, allowing them to stay connected as they move. Using this dual-mode model, network administrators can also reduce the number of WiFi access points needed to attain full campus coverage, thereby reducing maintenance costs.

## 1.3 MOBILE BROADBAND INTERNET USER SCENARIOS

The Internet continues to grow not only in number of subscribers and amount of traffic, but also in the types of traffic and the quantity of new applications. Fueled by growing broadband connectivity, the Internet is becoming richer in terms of multimedia applications and services. For example, in the past two years, there has been unprecedented growth in social networking applications, such as YouTube,* where users view over 100M video clips per day,10 and MySpace* which has over 100M users.11 Beyond complementing WiFi by extending affordable broadband connectivity outside of workplace, home, and public hotspots, WiMAX promises to deliver new usage models for subscribers. Connectivity from WiMAX and WiFi networks delivers exciting possibilities from real-time location awareness for social networks to real-time information sharing for mobile business productivity to extended education beyond the classroom. These new mobile Internet possibilities combined with existing user comfort levels with broadband and wireless networks are expected to reduce the barriers to user adoption for mobile broadband Internet.

## 1.4 THE WORLD INTERNET USAGE AND POPULATION STATISTICS

The considerable activity in the Internet and online markets across the region, the market in Asia continues to be dominated by the big players of north asia such as South Korea, Japan, Hong Kong and etc, at the same time the significant role star played by some of the south east asian countries such as in Singapore, Malaysia. In terms of sheer Internet user numbers, in China almost 300 million and India almost 95 million, maintain a real presence, despite their modest user penetration in figure 3-5. Especially china surpassed the US in 2008 to become the largest nation for the Internet users in the whole world and by end 2010 was showing no signs of slowing down even its become growing more faster than before.

In fact as highlighted in the Table 4, the Asian Internet market can be broken into three user penetration groups. The group includes the most highly penetrated markets in the world in terms of the users and subscribers. As well as they tend to have sophisticated and extensive broadband access facilities in place. Typically, in this research also we find that countries from this group are among the global leaders in broadband Internet as shown in Figure 2-7. The South Korea, with a user penetration of over 77%, leads the regional market with a broadband subscriber penetration of 32%. Closely following in second place is Hong Kong with a 75% user penetration and 28% broadband subscriber penetration.

The countries in the second band – roughly between 10% and 30% user penetration – are to be found in expansion mode when it comes to their Internet market segments. But there is a clear gap (almost 20%) to be bridged before they can be counted in the top grouping. China has jumped from 16% penetration in 2007 to 23% by mid-2009; however, its broadband penetration is still languishing around the 5% mark. Thailand and Vietnam (both with 26% user penetration) play lead roles in this middle grouping. In this research find that in the last few years the operators and governments have started to give main concern to increasing the Internet access and Internet speed in these kind of countries to make it more stronger such as Indonesia has moved strongly in the group, and the main challenges is to provide the Internet to cover the whole country, and it have 14% Internet user penetration and almost 33 million Internet users [18]. Furthermore, Malaysia also has moved strongly in the group, its reach almost 57% Internet user penetration and almost 15,300 million Internet users.

Nevertheless, In the 3ed group, those markets with a user Internet penetration of less than 10% user penetration, we tend to find those countries such as (India, Sri Lanka, Armenia, Afghanistan, Cambodia, Myanmar and etc), for whatever reason have not yet got their act together specially when it comes to Internet. And maybe, some of these countries are performing relatively well under difficult circumstances. The war-ravaged Afghanistan is managing 2% user penetration under extremely difficult circumstances. In this case there is the tiny fledgling nation of Timor Leste which has continued to struggle with political instability as it attempts to build its national infrastructure. As we shown in the table 4 at the bottom, this kind of countries such as Timor Leste has 0.2%Internet user penetration and almost 2 million Internet users, as well as Myanmar has 0.2%Internet user penetration and almost 120 million Internet users. However, we find a number of countries that are simply struggling with poor telecom infrastructure, communication and generally underdeveloped regulatory regimes. Included amongst these are Nepal, Turkmenistan, Cambodia, Tajikistan, Bangladesh and Myanmar. According to the report in [16] the Internet Usage and World Population Statistics are for June 2010 shown in the Figure 2.

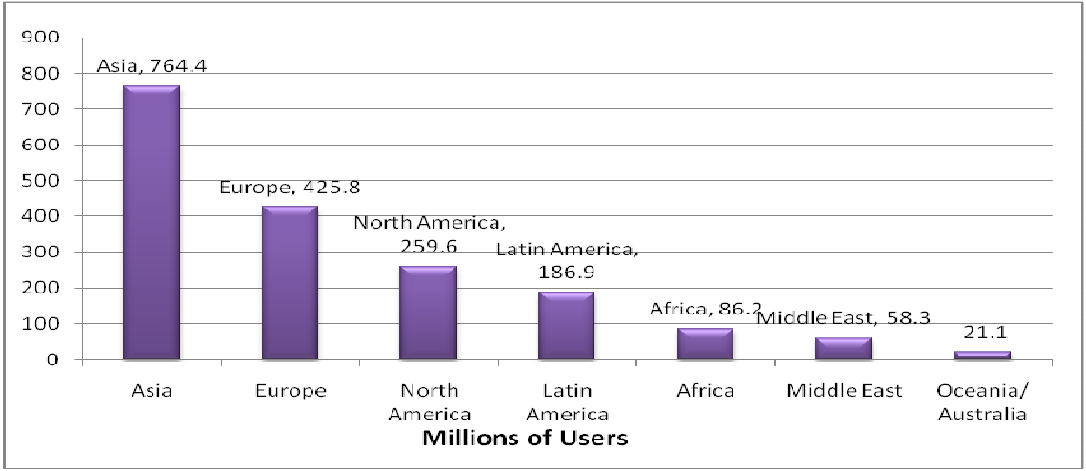

Figure 2. Internet Usage and World Population (Statistics the number of users in whole world in June 2010)

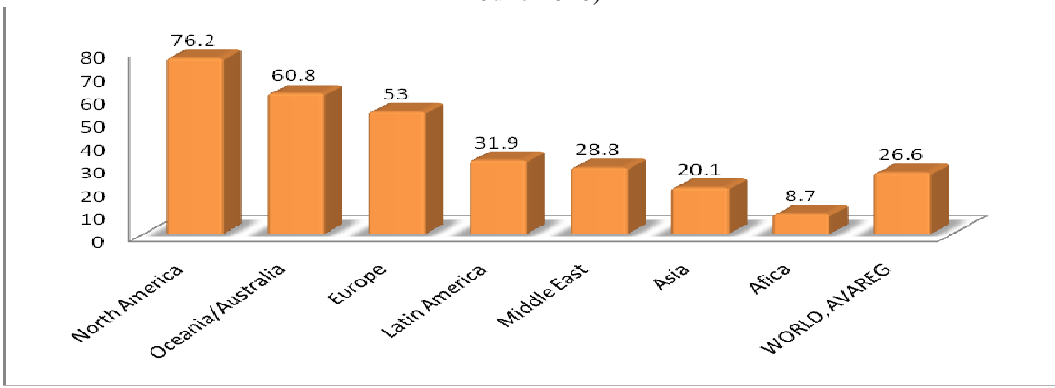

Figure 3. Penetration Rate

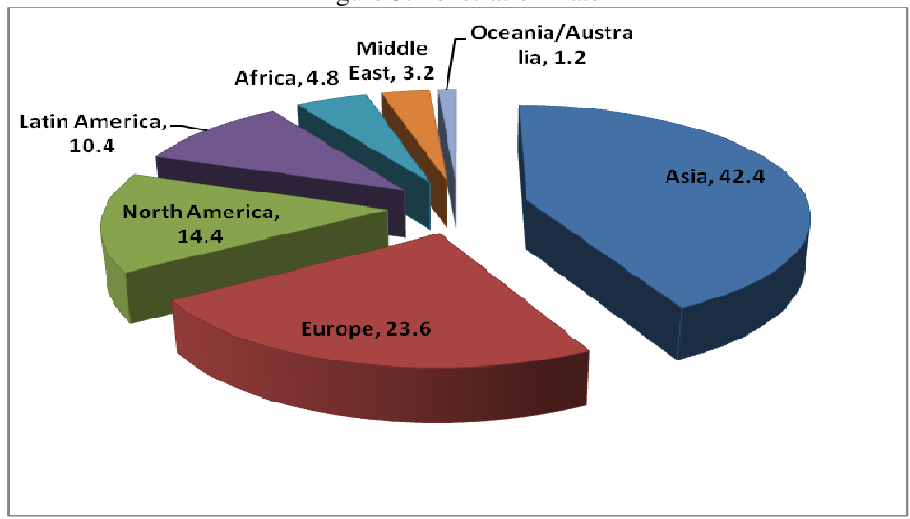

Figure 4.World Internet Users Distribution by World Regions -2009

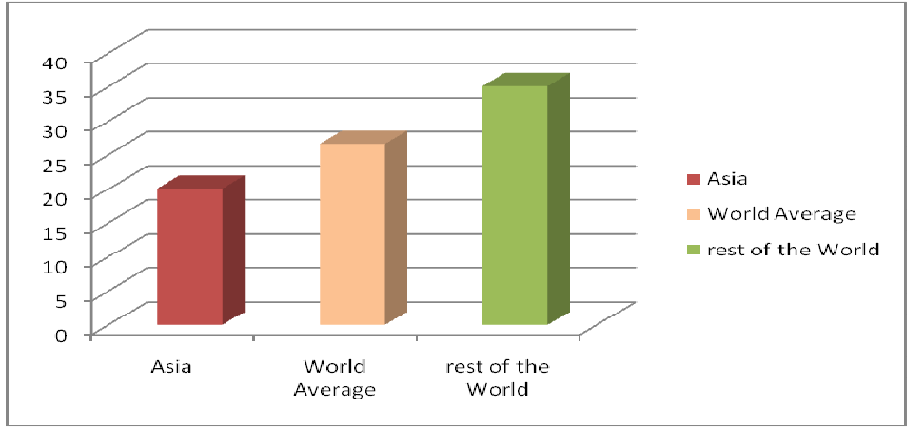

Figure 5.Internet Penetration in Asia 2009

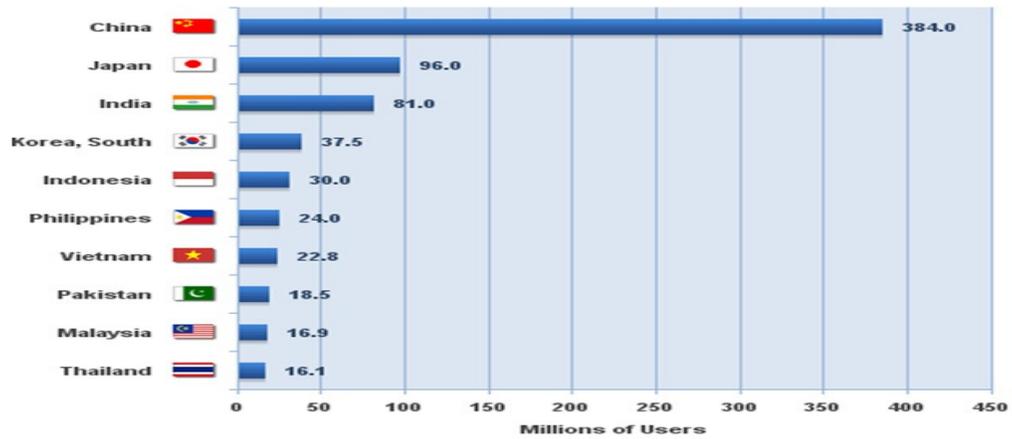

Figure 6. Internet in Asia 2009 (Top 10 Country)

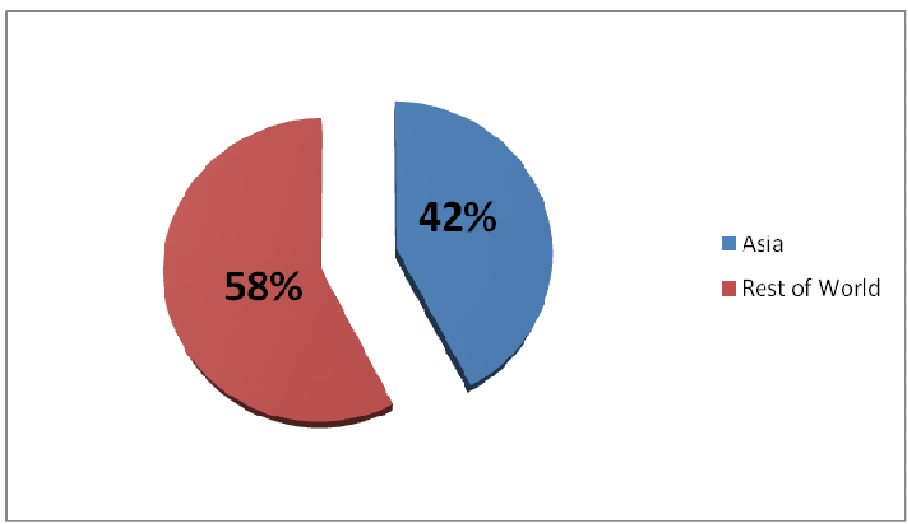

Figure 7. Internet Users in Asia (Asia Vs. World in 2009)

Table 4. Internet penetration and users in Asia by country – ranked by user penetration in 2009

| Country | Internet user penetration (%) | Internet users (thousand) |
|---|---|---|
| South Korea | 77% | 36,800 |
| Hong Kong | 75% | 5,250 |
| Singapore | 75% | 3,600 |
| Japan | 74% | 94,000 |
| Taiwan | 68% | 15,800 |
| Macau | 62% | 340 |
| Brunei Darussalam | 58% | 240 |
| Malaysia | 57% | 15,300 |
| Azerbaijan | 41% | 3,500 |
| Maldives | 27% | 80 |
| Georgia | 27% | 1,100 |
| Thailand | 26% | 17,500 |
| Vietnam | 26% | 22,500 |
| China | 23% | 298,000 |
| Kyrgyzstan | 17% | 900 |
| Kazakhstan | 16% | 2,500 |
| Indonesia | 14% | 33,000 |
| Mongolia | 13% | 400 |
| Pakistan | 11% | 19,000 |
| Uzbekistan | 10% | 2,800 |
| Tajikistan | 10% | 650 |
| India | 8% | 95,000 |
| Sri Lanka | 7% | 1,400 |
| Armenia | 6% | 200 |
| Bhutan | 6% | 40 |
| Philippines | 6% | 5,800 |
| Laos | 5% | 300 |
| Afghanistan | 2% | 500 |
| Nepal | 2% | 550 |
| Turkmenistan | 1% | 80 |
| Cambodia | 0.5% | 80 |
| Bangladesh | 0.4% | 600 |
| East Timor | 0.2% | 2 |
| Myanmar | 0.2% | 120 |

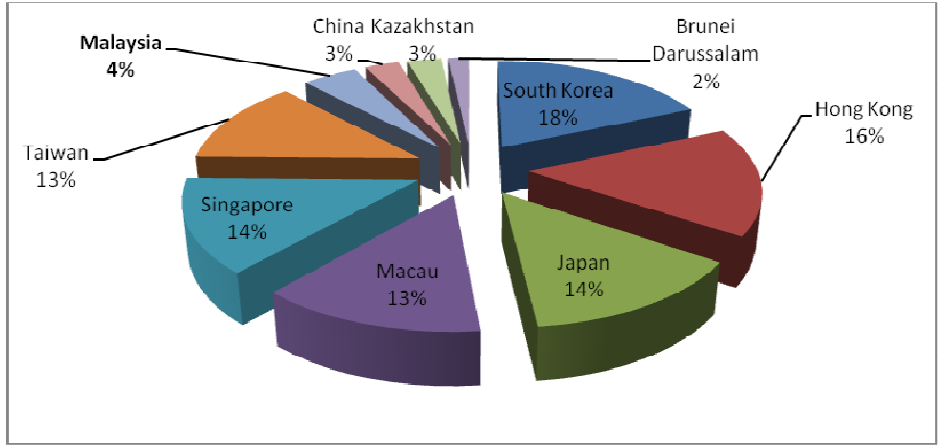

Figure 8. Top 10 broadband markets in Asia ranked by subscriber penetration (Country with the Broadband subscriber penetration in 2009

The Users are those accessing the Internet from school, university, cybercafes, and work accounts as well as from individual household or business accounts. For example, a work account is just one subscription but can have many users within that one subscription as shown in the Figure 8.With its widespread application of modern technologies such as fibre optics, wireless transmission, digitalisation and satellite services, Malaysia has one of the more advanced telecom networks in the developing world. Actually, the country has a national aim to see Malaysia ranked in the top developed country by year 2020. Especially Malaysia recently has worked harder to produce one of the more advanced telecom environments in the developing world. This aim need to building a new advanced telecom sector has had a strong links to national pride, certainly for a period in the 1990s the country busily promoted itself as a regional higher technology hub. So let us coming into 2009 virtually all almost 27 million people in Malaysia had a mobile phone service, which meant that Malaysia had the 2ed highest mobile penetration in Asia after Singapore, but to our knowledge Singapore had a small population comparing to Malaysia. The most significant is the growth in the mobile devices was continuing during 2009-2010, with a significant push into 3G services [13], and will be cover with 4G in the near future as well as some of the mobile devices is become provide the 4G services. The developmental effort in the communication has been led by a booming mobile market is almost more than 30 million subscribers and a penetration of 106% at the start of 2010, More recently, there has been a major push into the area of 3G services with around 21% of the total mobile subscriber base being 3G subscribers by end-2009 as shown in Figure 9-11.

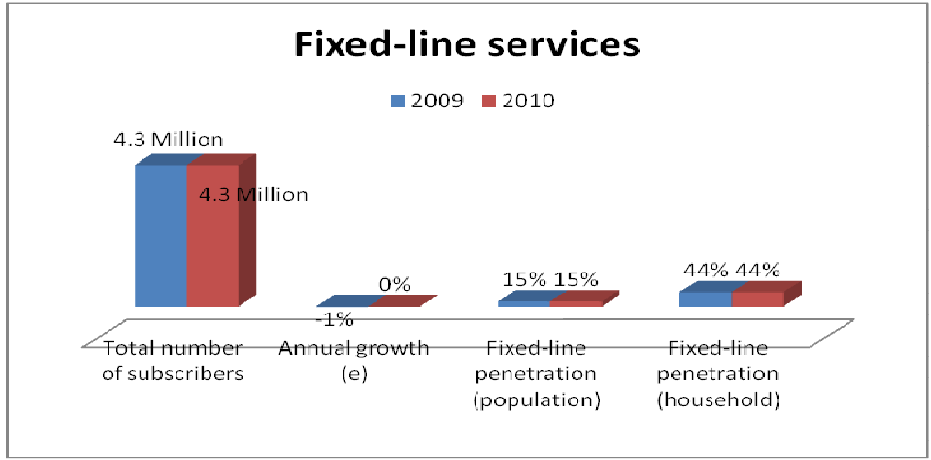

Figure 9. Fixed line Services

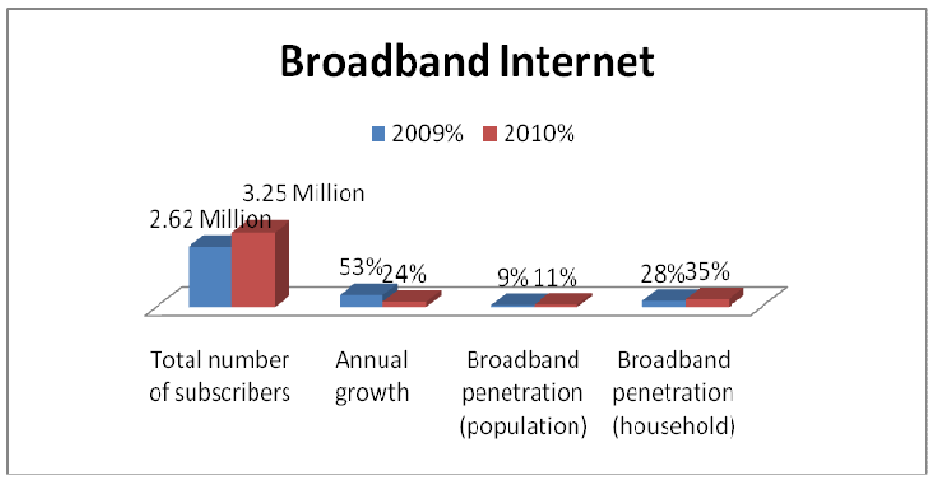

Figure 10. Shows the number of subscribes the broadband Internet

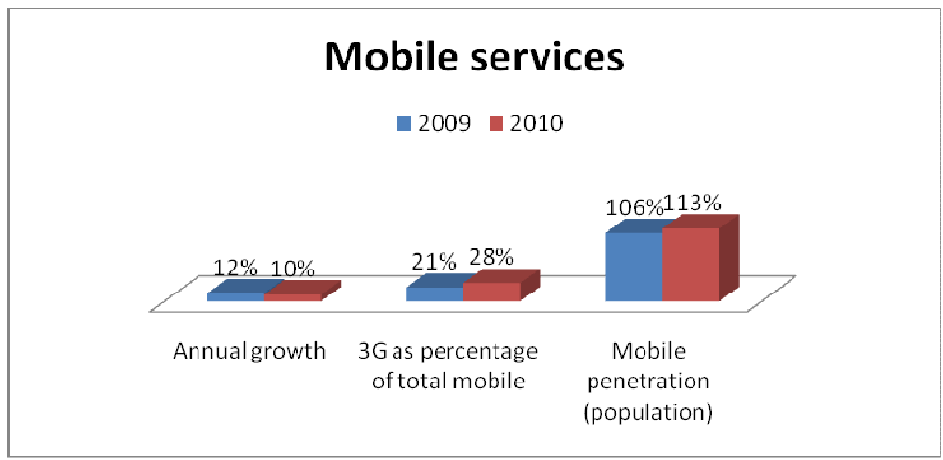

Figure 11: Shows the number of mobile services through 2009 -2010

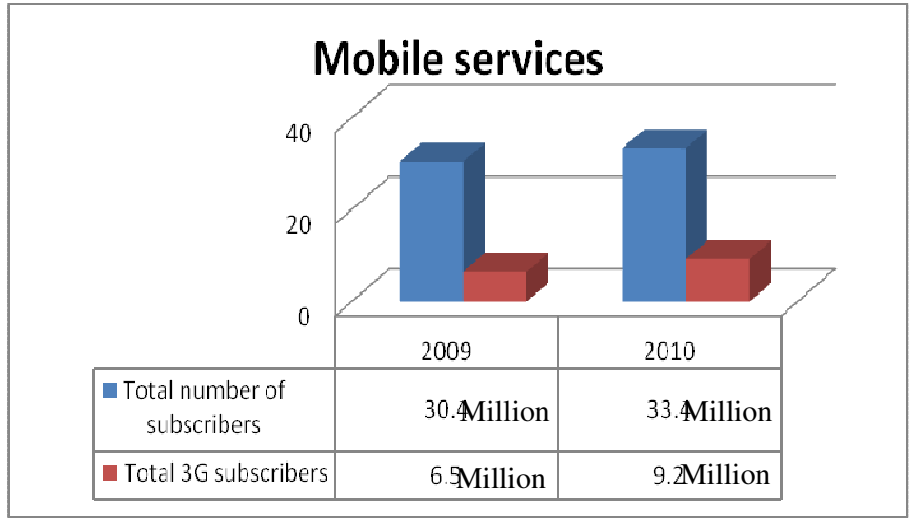

Figure 12: shows the total subscribers in Malaysia during 2009-2010

In additionally, according to the Figure 13, that the number of people using their mobile devices on a daily basis to access news and other information more than doubled during Jan 2008 within 10.8 million until Jan 2009 within 22.4 million. Furthermore, the use of laptops in many circumstances has given way to make it more powerful mobile devices.

According to the Figure 14, the number of people using their mobile devices to access online news, entertainment through mobile such as YouTube, Facbook and other information on a daily basis is rising sharply. Whether daily, weekly or at some point during a given month were just less than 63.2 million people in January 2009. In Figure 15 shows some 8.2 million users downloaded maps, making that the most popular downloadable application, as well as almost 14.1 million users utilized their SMS based news and information access for search, making that the most favoured use.

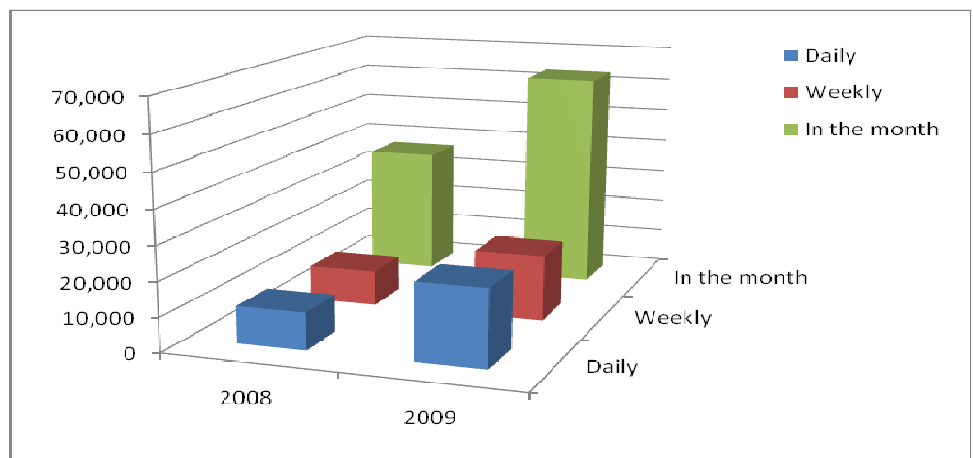

Figure 13.Frequency of mobile Internet access from 2008 to 2009

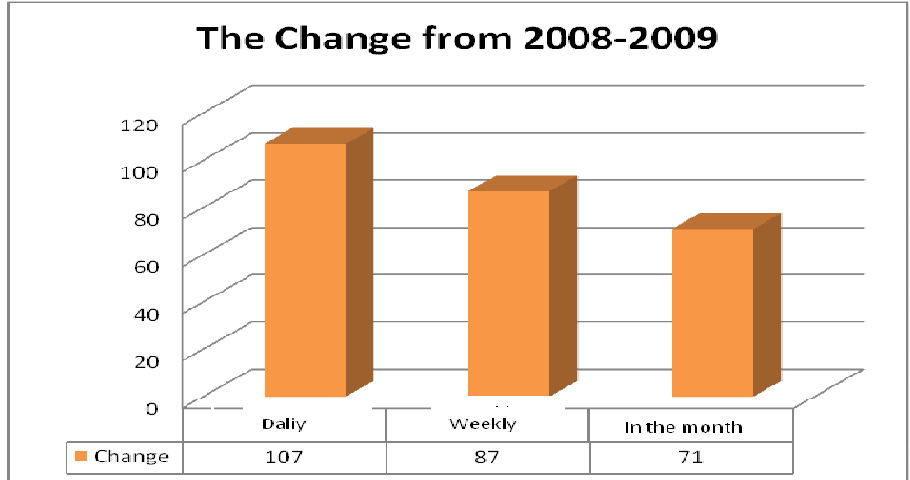

Figure 13. The Frequency of mobile Internet Access three month average Ending January 2008 and January 2009, Mobile phone users

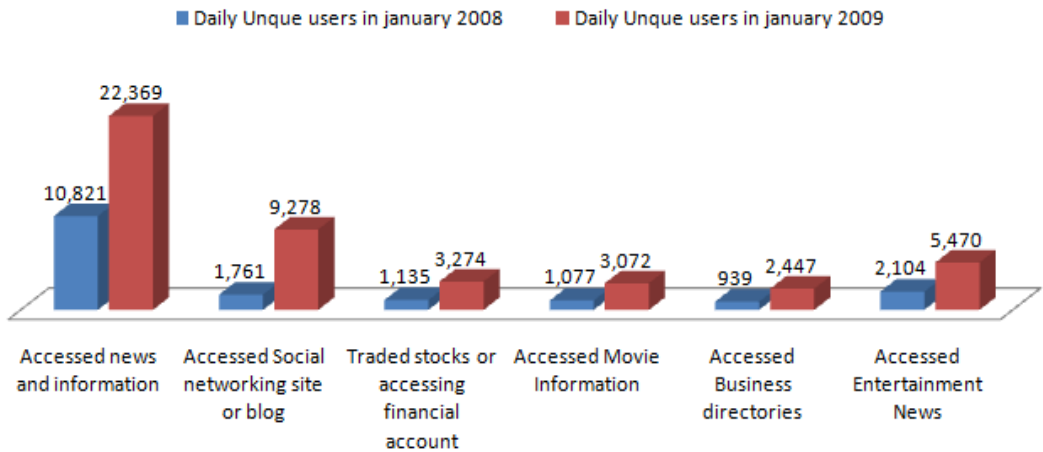

Figure 15. Shows the growing categories of daily mobile web access during three month

Regarding to the report in [17] and [18] more than 54 million analog mobile television enabled cellular phones in 2011 and almost 300 million analog mobile TV users in 2013 as explained that in Figure 16 also according the press release and free downloadable white paper [21] [22].

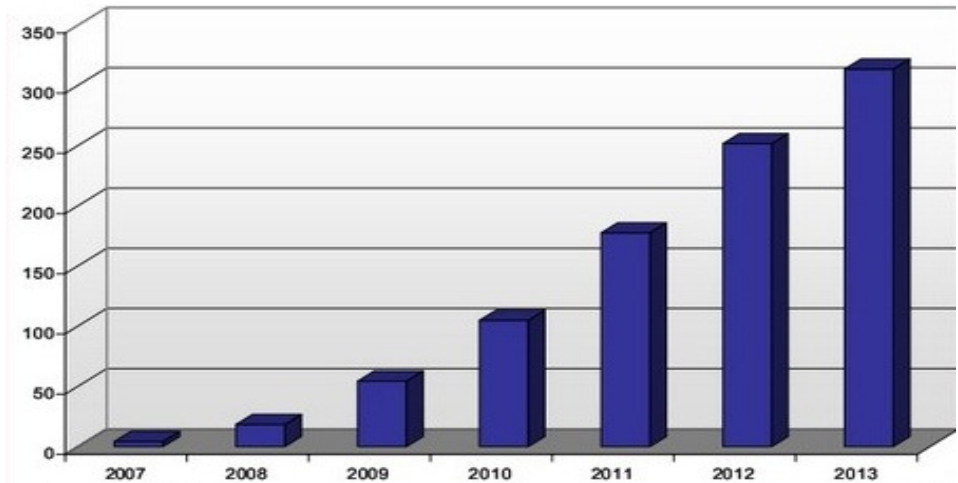

Figure 16. Worldwide forecast of analog free to air mobile TV

In the analog mobile TV, the world's most widely available option for mobile TV. Actually, Based on the findings on research conducted in the first quarter of 2009 through online surveys filled out by consumers in Brazil, Colombia, Indonesia and Turkey. Based on the findings from a research conducted in 2009, more than 100 mobile TV operators worldwide offering services ranging from 3 to 20 $ monthly. The telegent, manufactures chipsets for analog mobile TV has shipped more than 40million since 2007. The cost to add analog mobile TV to a device is typically less than $10. Finally, regarding to the infonetics research forecasts that more than" 397 million cellular video phones, creating a market worth tens of billions of dollars will be sold in 2013" as shown in Figure 17 [25], at the end of 2008 there were 41 million video phone subscribers, its means that number will increase more than 10 times by 2013.

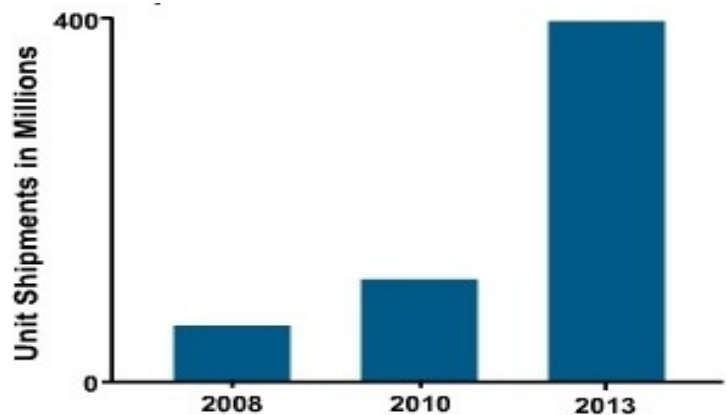

Figure 17. Strong growth in worldwide mobile video phone shipments continues into 2013.

According to the mobile video equipment, Phones, Services and Subscribers, will become the dominant mobile television technology. In this time, its almost more than 1 million subscribers in china already utilize china multimedia broadcasting. Furthermore, The Asia has the greatest volume of mobile video sales and that is certainly no surprise, as well as the sports, such as (cricket, soccer and automobile racing) is the primary driver of mobile video use that also isn't a surprise, and there many of the video are providing to make the mobile client enjoy with their time at anytime and anywhere.

# CONCLUSIONS

Mobile devices are becoming the most important communication tool in the world which is evident from the statistical information about the Internet, Communications and mobile devices etc. This has led to an increased demand for the development, communication and computational powers of many of the mobile wireless subscribers/mobile devices such as laptops, PDAs (Personal Digital Assistants), smart phones and notebook. These techniques are utilized to obtain a video on demand service with higher resolution and quality, especially when the subject requires the capacity to move from place to place.

## ACKNOWLEDGEMENTS

Special thank and recognition go to my advisor, Associate Professor. Dr. Putra Sumari, who guided me through this study, inspired and motivated me. Last but not least, the authors would like to thank the School of Computer Science, Universiti Sains Malaysia (USM) for supporting this study.

## Authors

Saleh Ali K. Al-Omari has obtained his Bachelor degree in Computer Science from Jerash University, Jordan in 2004-2005 and Master degree in Computer Science from Universiti Sains Malaysia, Penang, Malaysia in 2007. Currently, He is a PhD candidate at the School of Computer Science, Universiti Sains Malaysia. He is the candidate of the Multimedia Computing Research Group, School of Computer Science, USM. He is a member and reviewer of several international journals and conferences (IEICE, JDCTA, IEEE, IACSIT, AIRCC, etc). His main research area interest are in includes Peer to Peer Media Streaming, Video on Demand over Ad Hoc Networks, MANETs, and Multimedia Networking, Mobility.

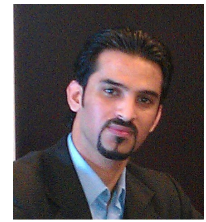

Putra Sumari obtained his MSc and PhD in 1997 and 2000 from Liverpool University, England. Currently, he is a lecturer at the School of Computer Science, Universiti Sains Malaysia, Penang. He is the head of the Multimedia Computing Research Group, School of Computer Science, USM. Member of ACM and IEEE, Program Committee and reviewer of several International Conference on Information and Communication Technology (ICT), Committee of Malaysian ISO Standard Working Group on Software Engineering Practice, Chairman of Industrial Training Program School of Computer Science USM, Advisor of Master in Multimedia Education Program, UPSI, Perak.

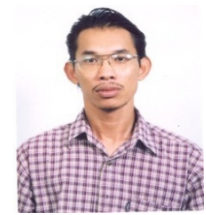